\documentclass[12pt,dvips]{article}
\usepackage{graphicx}
\usepackage{epsfig}

\newcommand{\resection}[1]{\setcounter{equation}{0}\section{#1}}

\renewcommand{\theequation}{\thesection.\arabic{equation}}
\def\be{\begin{equation}}
\def\ee{\end{equation}}
\def\bea{\begin{eqnarray}}
\def\eea{\end{eqnarray}} 
\begin{document}
\oddsidemargin 5mm
\setcounter{page}{0}
\newpage     
\setcounter{page}{0}
\begin{titlepage}
\begin{center}
{\large {\bf Special Theory of Relativity through the Doppler Effect}}\\
\vspace{1.5cm}
{\bf M. Moriconi} \footnote{\tt{email:moriconi@if.uff.br}}\\
{\em Departamento de F\'\i sica}\\
{\em Universidade Federal Fluminense}\\
{\em Av. Litor\^anea s/n, Boa Viagem - CEP 24210-340}\\
{\em Niter\'oi, Rio de Janeiro, Brazil}\\
\vspace{0.8cm}
\end{center}
\renewcommand{\thefootnote}{\arabic{footnote}}
\vspace{6mm}

\begin{abstract}
\noindent
We present the special theory of relativity taking the Doppler effect as the starting point, and derive several of its main effects, such as time dilation, length contraction, addition of velocities, and the mass-energy relation, and assuming energy and momentum conservation, we discuss how to introduce the 4-momentum in a natural way. We also use the Doppler effect to explain the ``twin paradox", and its version on a cylinder. As a by-product we discuss Bell's spaceship paradox, and the Lorentz transformation for arbitrary velocities in one dimension.
\vspace{3cm}

%PACS number(s): 11.55, 05.50+q  
\end{abstract}
\vspace{5mm}
\end{titlepage}

\newpage
%\baselineskip=24pt
%\pagebreak
\setcounter{footnote}{0}
\renewcommand{\thefootnote}{\arabic{footnote}}

\resection{Introduction} 

During 2005 we celebrated Einstein's {\em annus mirabilis}, a year in which the course of physics changed in a profound way. There were several events celebrating Einstein's contributions, and this gave us the opportunity to go through some of the fundamentals of physics, following the path Einstein laid out for us, and rethinking ways to present some of his ideas, in particular the special theory of relativity (STR) \cite{Principle}.

In this article we present an elementary approach to the STR, taking the Doppler effect as starting point . We will treat only one dimensional motion, for simplicity, but this is certainly not a limiting asumption, since one can easily generalize to arbitrary motions by combining Lorentz boosts and rotations. Initially we derive the Doppler effect formula {\em directly} from the two postulates of the STR, namely

\begin{itemize}

\item The laws of physics are the same in all inertial frames of reference

\item The velocity of light is the same for all inertial observers

\end{itemize}
As a matter of fact one should consider only one postulate, since the principle of the constancy of the speed of light is a consequence of the first postulate, as Einstein points out in the $E=Mc^2$ paper (see, for example, \cite{Principle}): all one needs to do is to qualify {\em what} are the laws of physics that are supposed to be valid in all frames of reference. If one assumes Newton's laws, one obtains classical kinematics. If one takes Maxwell's equations, then one is naturally led to a new view of space and time.

Once we have derived the relativistic Doppler formula, we apply it to several different thought experiments, which will allow us to derive all of relativistic kinematics: time dilation, length contraction, the addition of velocities, the Lorentz transformations, and the mass-energy relation. We also present a solution of the twin paradox through the Doppler effect, and its version on the cylinder. We discuss Bell's two spaceships paradox, and the Lorentz transformation for arbitrary velocities in one dimension. In deriving the mass-energy relation we show, as a by-product, that the energy of radiation suffers a Doppler shift too, without having to resort to the energy-frequency relation of elementary quantum mechanics. This will allow us to introduce the ``four-momentum" of a particle in a quite natural way. By now there is a great number of good books on the special theory of relativity, such as French's or Taylor and Wheeler's \cite{Books}, which usually rely on space-time diagrams. Our approach is based solely on the Doppler effect, and we believe it can be used as a tool to understand the essence of relativity.

\resection{The Doppler Effect}

Consider a light source moving along the $x$ axis with velocity $v$ and emitting light of frequency $\nu_0$, as measured in the inertial frame of the source. Let us call Isabella an observer towards which the light source is moving to, and Marianna an observer from which the light source is moving away. Both, Isabella and Marianna, are at rest in the lab frame. The frequency observed by any one of them should be a function of $\nu_0$, $v$ (the velocity of the source, as observed by them) and $c$. Elementary dimensional analysis gives $\nu(v)=\nu_0 f(\beta)$, where $\beta=v/c$, and $f(\beta)$ is an unknown function. From the point of view of Isabella, if a pulse is emitted at instant $t_1$ and a second pulse at $t_2$, she will then observe a wavelength given by $(c-v) \Delta t$, where $\Delta t = t_2-t_1$, and the observed frequency is therefore $\nu(v)$. From the point of view of Marianna, we must replace $v \to -v$, since the light source is moving away from her. According to the postulate of the constancy of the velocity of light, we have:
\bea
&&(c-v)\Delta t \nu(v) =c  \ , \nonumber \\
&&(c+v)\Delta t \nu(-v)=c \ . \
\eea
Note that we would have obtained the same equations if the observers were moving, instead of the source, due to the first postulate (all that matters is their relative motion). Dividing the first equation by the second, we obtain
\be
\frac{f(\beta)}{f(-\beta)}=\frac{1+\beta}{1-\beta} \label{f}
\ee
We can now use the first postulate: if the source {\em and} the observer move at the same velocity $v$, she will clearly observe the same frequency $\nu_0$ emitted by the light source, since they are in the same inertial frame. But according to our formula for $\nu(v)$, this is also equal to $\nu_0f(\beta)f(-\beta)$: the light observed by someone at rest in the lab frame, between the source and the travelling observer, has frequency $\nu=\nu_0 f(\beta)$, and this light will be seeing by the moving observer to have the frequency $\nu'=\nu f(-\beta)$. Therefore $f(\beta)f(-\beta)=1$. We can use this in (\ref{f}), to obtain
\be
f(\beta)^2=\frac{1+\beta}{1-\beta} \Rightarrow f(\beta)=\left( \frac{1+\beta}{1-\beta}\right)^{1/2}
\ee
This gives us the Doppler effect formula:
\begin{equation}
\nu(v)=\nu_0\left(\frac{1+\beta}{1-\beta}\right)^{1/2}
\end{equation}
There is another elementary derivation of this formula, directly from the postulates of the STR, which we leave for appendix 1, in order not to interrupt the flow of the article. We move on now, to apply this formula to different set-ups and obtain the main consequences of relativistic kinematics.

\resection{The Addition of Velocities}

The addition of velocities may be easily derived with the aid of the Doppler effect formula. Let us suppose Isabella moves with velocity $v$ from left to right, and Marianna, who owns a light source of frequency $\nu_0$, moves with velocity $u$ from right to left. For simplicity assume they are approaching each other. Both velocities are measured relative to their father's laboratory frame, who is in between the two. He will measure a frequency $\nu_{f}=\nu_0((1+u/c)/(1-u/c))^{1/2}$ for Marianna's light. This means that Isabella should observe $\nu=\nu_{f}((1+v/c)/(1-v/c))^{1/2}$. But since we can write, in Isabella's frame, that $\nu=\nu_0((1+w/c)/(1-w/c))^{1/2}$, where $w$ is the relative velocity, we have
\be
\left(\frac{1-u/c}{1+u/c}\right)^{1/2} \left(\frac{1-v/c}{1+v/c}\right)^{1/2} = \left(\frac{1-w/c}{1+w/c}\right)^{1/2} 
\ee
from which we readily derive
\begin{equation}
  w=\frac{u+v}{1+uv/c^2}
\end{equation}
We will use this equation again when deriving the mass-energy relation.
\resection{Time Dilation}

We can use the Doppler effect to derive the time dilation effect. While deriving the Doppler effect, we found that
\be
(c-v) \Delta t \nu_0 \left(\frac{1+\beta}{1-\beta}\right)^{1/2} = c \Delta \tau \nu_0 \ ,
\ee
where the right-hand side is the velocity of light in the light-source frame, that is, the product of the wavelength and frequency as measured by an observer that moves along with the light source. $\Delta \tau$ is the time interval between two pulses in that frame. One obtains immediately
\begin{equation}
  \Delta t = \frac{\Delta \tau}{(1-\beta^2)^{1/2}} \ ,
\end{equation}
which is the time dilation effect. Notice that in most elementary presentations of relativity one derives the time dilation {\em first} and then uses it to derive the Doppler effect expression.
\resection{Length Contraction}

Suppose now that Marianna takes off from rest, till she reaches velocity $v$($<c$, of course) from the origin of Isabella's inertial frame $S_0$, goes to a point $x_0$ at distance $L_0$, as measured by Isabella, and comes back to the origin with the velocity $-v$, and stops at the origin. Let us assume that there is a far away light source in the same line that Marianna is moving, emitting light pulses with frequency $\nu_0$, and that Marianna's acceleration and deceleration times are short, compared to the total traveling time, even though this is not crucial in the derivation, as we will see. Initially we should note that each light pulse that passes through Marianna will necessarily reach Isabella before she comes back to the origin, since $v<c$. So, if even after taking note of a given pulse, Marianna immediately decides to go back to the origin, that pulse will reach Isabella {\em before} Marianna reaches her. Moreover, each pulse that reaches Isabella necessarily passed through Marianna. Therefore the number of pulses counted by Marianna is exactly the same number of pulses counted by Isabella. Let us count how many pulses Isabella observes: the total travel time measured by her is $2L_0/v$, and so the number of pulses she counts is given by $N_I=\nu_0 2L_0/v$. In order to compute the number of pulses Marianna observes, we must use the Doppler formula. On her way to  to $x_0$ she observes $N_+=\nu_0 ((1+\beta)/(1-\beta))^{1/2} L'/v$. Here we are considering that she observes the point $x_0$ at an unknown distance $L'$, and that this point moves toward her with velocity $v$. On her way back she will observe 
$N_-=\nu_0 ((1-\beta)/(1+\beta))^{1/2} L'/v$. Note that on her way back she observes the same distance $L'$ to the origin. The fact that $N_M=N_++N_-=N_I$ gives
\be
\frac{2L'/v}{(1-\beta^2)^{1/2}}=2L_0/v \Rightarrow 
L'=L_0(1-\beta^2)^{1/2} \ , \label{LF}
\ee
which is the length contraction effect \footnote{If one feels uneasy with the round trip in this argument, we may provide the following variation: when Marianna reaches $x_0$, she has observed $\nu_0((1+\beta)/(1-\beta))^{1/2} L'/v$ pulses, whereas Isabella has observed $\nu_0 L_0/v$ pulses. The difference between these two numbers should be exactly the number of pulses between $x_0$ and Isabella: these are the pulses that have crossed Marianna, but not Isabella. This number is given by $\nu_0 L_0/c$. Therefore we have
$\nu_0((1+\beta)/(1-\beta))^{1/2} L'/v=\nu_0 L_0/v+ \nu_0 L_0/c$, which gives the result (\ref{LF})  again.}(FitzGerald-Lorentz contraction).

\resection{The Twin Paradox}

Incidentally this counting procedure also solves the famous ``twin paradox", or in  this case, the ``sister's paradox": two sisters, Isabella and Marianna, are separated. Isabella  stays in an inertial frame and Marianna goes on a round trip just like the one we described. Each one of them sees the other's clock run slower than her own. When they meet again who is right? According to Isabella, there where $2T\nu_0$ pulses, and according to Marianna there were $T'((1+\beta)/(1-\beta))^{1/2} \nu_0 + T'((1-\beta)/(1+\beta))^{1/2} \nu_0$. Since they observe the same number of pulses, we must have  $T'=T(1-\beta^2)^{1/2}$, which means that Marianna's clock indeed ran slower than Isabella's.

It is also possible to use this method to derive the time difference for an arbitrary motion. The counting argument requires only that the velocity of the traveler is always smaller than the velocity of the light. Therefore, let us consider an arbitrary velocity function for the distant light source as a function of time $v(\tau)$, where $\tau$ is Marianna's proper time. Between instants $\tau$ and $\tau+d\tau$, Marianna observes $dN(\tau)=d\tau \left((1+\beta)/(1-\beta)\right)^{1/2}\nu_0$ pulses. During the whole trip she will have observed
\be
N=\int_0^{T'} d\tau \left(\frac{1+\beta}{1-\beta}\right)^{1/2}\nu_0
\ee 
pulses, which should be equal to Isabella's counting, $T \nu_0$. This gives the relation
\be
T=\int_0^{T'} d\tau \left(\frac{1+\beta}{1-\beta}\right)^{1/2}=\int_0^{T'} d\tau \frac{1+\beta}{(1-\beta^2)^{1/2}}
\ee
In the case the observer returns to the origin, the second term on the left-hand side vanishes
\be
\int_0^{T'} d\tau \frac{\beta}{(1-\beta^2)^{1/2}}=0
\ee
since this is the total distance covered by the traveling observer. 
We will show this more formally in section 10, where we discuss the Lorentz transformations for arbitrary velocities.
Therefore we obtain the more familiar looking result
\be
T=\int_0^{T'} d\tau \frac{1}{(1-\beta^2)^{1/2}}
\ee
Since $(1-\beta^2)^{1/2}<1$, we see that $T>T'$, always.

\resection{The Twin Paradox on a Cylinder}

There is an interesting variation of the twin paradox, where one considers the space to be a cylinder, that is, the spacial dimension is compact. The two paradoxes seem very similar, but there is one crucial difference. In the previous paradox, ultimately, the explanation was due to the fact that, whereas one of the observers is always receiving pulses at the same frequency, the other observes blue-shifted and then red-shifted signals, due to the fact that she changed her velocity in order to go back to the origin. This is equivalent to the more usual explanation where it is pointed out that the situation is not really symmetric, since one of the observers had to accelerate, whereas the other did not, breaking the symmetry between the two.

In the case of a cylinder one can not use this ``breaking of the symmetry" explanation. This is so because there is more than one geodesic path that takes the traveling observer back to the origin, which is always an inertial frame of reference. Therefore there is no apparent asymmetry in this case.

We can use our counting method, similarly to what we have done in the classic twin paradox case. Suppose Isabella and Marianna are located at the origin, and that Marianna goes on a round trip, winding around the cylinder, which has radius $R$. Let us consider a light source at distance $L_1$ from the origin, measured in the clockwise direction, which sends light pulses with frequency $\nu_0$. This source has been sending pulses for a long time, and we can assume that at the beginning of Marianna's trip, both observers start counting pulses. We will also assume that the source is sending pulses towards the origin along the path of length $L_1$ until Marianna reaches it, and then the light source will send signals along the path of length $2\pi R -L_1$. Let us count the number of pulses in the two frames of reference.

According to Marianna, in the first leg of her trip she observes pulses at a higher rate, given by
\be
N_1^{m}=\tau_1 \left(\frac{1+\beta}{1-\beta}\right)^{1/2}\nu_0
\ee
where $\tau_1$ is the time she takes to reach the light source. On her way back to the origin, light is observed at a lower rate, and the number of pulses is simply given by
\be
N_2^{m}=\tau_2 \left(\frac{1-\beta}{1+\beta}\right)^{1/2}\nu_0
\ee
where now $\tau_2$ is the time Marianna takes to go back to the origin. The total number of pulses that Marianna has counted is
\be
N^{m}=\left(\tau_1 \left(\frac{1+\beta}{1-\beta}\right)^{1/2}+\tau_2 \left(\frac{1-\beta}{1+\beta}\right)^{1/2}\right)\nu_0 \label{nm}
\ee

According to Isabella we have the following. Let $T_1$ ($T_2$) be the time Marianna takes to reach the light source (reach the origin from the light source). During the first leg of Marianna's trip, Isabella observes $T_1 \nu_0$, but in order that their counting matches, that is, they count the same physical pulses emitted from the source, we have to add all the pulses that are between Isabella and the source, at the time Marianna reached it, which is equal to $L_1 \nu_0/c $. This gives
\be
N_1^{i}= (T_1 + \frac{L_1}{c})\nu_0
\ee
During the second leg of Marianna's trip, each pulse that she counts crosses her and reaches Isabella {\em before} she reaches the origin, and each pulse counted by Isabella necessarily crosses Marianna. We are assuming that Marianna takes a time $T_2$ to arrive at the origin, but since the light source only starts sending pulses {\em after} Marianna crosses it, Isabella will have to wait a time equal to $L_2/c$ to start receiving pulses. This gives
\be
N_2^{I}=(T_2-\frac{L_2}{c})\nu_0
\ee
and the total number of pulses counted by Isabella is given by
\be
N^{i}=\left(T_1 + \frac{L_1}{c}+T_2-\frac{L_2}{c}\right)\nu_0 \label{ni}
\ee
Picking $L_1=L_2$ gives, by symmetry $\tau_1=\tau_2=\tau/2$ and $T_1=T_2=T/2$, where $\tau$ and $T$ are the times of how long the round trip takes according to each observer. Substituting these in (\ref{ni}) and (\ref{nm}) we obtain
\be
\tau=T(1-\beta^2)^{1/2} \label{taut}
\ee
which is the same conclusion as in the classic twin paradox \footnote{If we take $L_1 \neq L_2$, equation (\ref{nm}) changes to $N^m=\tau(1+\beta(L_1-L_2)/L)/(1-\beta^2)^{1/2}$, since, by symmetry, $\tau_1/\tau_2=L_1/L_2$, and equation (\ref{ni}) becomes $N^i=T(1+\beta(L_1-L_2)/L)$, where we used that $T=(L_1+L_2)/v$, and we get the same result as (\ref{taut}).}.

In this analysis we see how useful the ``counting approach" is. Whereas in the usual solutions of the twin paradox in a cylinder one needs to do quite involved arguments, in this case one hardly sees any difference between the analysis of the classical twin paradox and this one.

The reader may still feel a little uneasy, in the search for the hidden asymmetry between the two systems of reference. In this case one can convince oneself by noticing the following: one of the observers is always receiving pulses at frequency $\nu_0$, whereas the other one sees a higher frequency during one leg and a lower frequency during the other. We can see, then, that ultimately, the reason for the asymmetry between the two observers is that there's a preferred global inertial frame \cite{Dray, Weeks, cylinder}.

\resection{Bell's Two Spaceship Paradox}

J. Bell has presented a quite interesting paradox in the special theory of relativity, which has caused (and apparently it still causes!) heated debate among physicists. The paradox is explained in his book \cite{Bell}, and we summarize it here. Suppose two spaceships, $I$ and $M$, are at rest in a given inertial frame, separated by a distance $L_0$, as measured in the lab frame, and suppose that an observer, located exactly at the midpoint joining the two spaceships, emits a light signal towards them, which serves as a signal for them to start their trip. Each spaceship is equipped with a program that tells the crew how to accelerate it, and the programs in $I$ and $M$ are exactly the same. The first question is: will the distance between the two spaceships be Lorentz-contracted according to the lab frame? The answer to this question is, clearly, `no´. The second question is, and here is the apparent paradox: if there was a thin thread joining the two spaceships, then, since the measurement in the lab frame gives $L_0$ for their separation, then how come {\em the thread should be Lorentz-contracted} according to, say, the observers in $I$? In other words, if the measurement in the lab frame gives $L_0$, then the length of the thread as measured by the spaceships should be {\em bigger} than $L_0$, in order to compensate the Lorentz contraction. There is something clearly wrong in this conclusion.

To answer the second question consider the following: if we observe, in the lab frame, that their distance is $L_0$, then in $I$'s frame their distance should be {\em bigger} than $L_0$, but since $I$ and $M$ have exactly the same velocity programs, one could be lead to think that their distance should be constant {\em according to them}. This is where the solution of the paradox stems, and can be easily understood by recalling one of the fundamental aspects of relativity: simultaneity is not an absolute concept. The lab measurement of the positions of the two spaceships occurs at a given time $t$ in the lab frame, but this is {\em not} the case in the spaceship's frames. Let us denote the measurements of the positions of the $I$ and $M$ spaceships, in the lab frame, by events ${\cal E}_I$ and ${\cal E}_M$. The corresponding events, as seen by the $I$ frame are denoted by ${\cal E}^{1}_I$ and ${\cal E}^{1}_M$. At the moment that a measurement is performed at $I$, what are the locations in space-time of events ${\cal E}^{1}_I$ and ${\cal E}^{1}_M$?  We postpone this to section 11, where we find the Lorentz tranformation for an arbitrary motion.

\resection{The Lorentz Transformation}

We can also derive the Lorentz transformation of coordinates, using a variation of these counting arguments. Let us suppose that in the inertial frame $S_0$ there is an observer, Isabella, at the origin of $S_0$. A second inertial frame, $S$, where Marianna is, moves along the $x$ axis with velocity $v$. A given physical event ${\cal E}$ will have coordinates $(x,t)$ according to Isabella, and coordinates $(x',t')$ according to Marianna. The relation between these coordinates is given by the Lorentz transformation.

Let us suppose that there is a light source that has been emitting light for a long time, and that it is located very far away, in such a way that there are light pulses passing through the origin of $S_0$ since before $t=0$ (which we assume is when both origins coincide). We will take as the physical event ${\cal E}$, the crossing of a light pulse at $x$ at time $t$. When this happens, Isabella has already observed $t \nu_0$ pulses, and there are $\nu_0 x/c$ pulses between her and $x$, totaling $N_1=(t+x/c)\nu_0$ pulses. According to Marianna, though, she has counted $t'((1+\beta)/(1-\beta))^{1/2}\nu_0$ pulses, and there are $((1+\beta)/(1-\beta))^{1/2}\nu_0 x'/c$ pulses between her and the physical location of ${\cal E}$, totaling $N_2=t'((1+\beta)/(1-\beta))^{1/2}\nu_0+((1+\beta/(1-\beta))^{1/2}\nu_0 x'/c$ pulses \footnote{This can be derived using the principle of relativity: since the total number of pulses Isabella measured is given by $N_I=(t+x/c)\nu_0$, then the total number of pulses measure by Marianna {\em must be} $N_M=(t'+x'/c)\nu$, which is the expression we found.}. The two numbers $N_1$ and $N_2$ have to be equal, which gives the following equation
\be
t'\left(\frac{1+\beta}{1-\beta}\right)^{1/2}+\frac{x'}{c}\left(\frac{1+\beta}{1-\beta}\right)^{1/2}=t+\frac{x}{c} \ .
\label{counting}
\ee
Note that his equation is the statement of the fact that the phase of a plane wave is a relativistic invariant.
Before we proceed, we can show that (\ref{counting}) implies the end of absolute simultaneity. In order to show that, consider two physical events ${\cal E}_1$ and ${\cal E}_2$, separated by a distance $L_0$, and that occur at the same time $t$ according to Isabella. Assuming that Marianna also observes simultaneous physical events (that is, that $\Delta t' =0$ for her too), we would have that the distance between ${\cal E}_1$ and ${\cal E}_2$ is given by $\Delta x = L_0 (1-\beta^2)^{1/2}$, and so (\ref{counting}) implies
\be
\frac{L_0}{c}(1+\beta)=\frac{L_0}{c}
\ee
which is true only if $\beta=0$: Isabella and Marianna must be at rest in relation to each other. This is the end of absolute simultaneity.

In order to find the Lorentz transformation for $x$ and $t$ we still need one more equation. Suppose, then, that another event ${\cal E}'$ takes place at $(-x,t)$ according to Isabella, and that Marianna moves with velocity $-v$. What are the coordinates of ${\cal E}'$ according to Marianna? They are simply given by $(-x',t')$, since all we have done was to perform a parity transformation \footnote{We are assuming that the STR coordinates do not break parity.}, and so we can write
\be
t'\left(\frac{1-\beta}{1+\beta}\right)^{1/2}-\frac{x'}{c}\left(\frac{1-\beta}{1+\beta}\right)^{1/2}=t-\frac{x}{c} \ .
\label{counting-}
\ee
Adding and subtracting (\ref{counting}) and (\ref{counting-}) we obtain the Lorentz transformation
\be
x=\frac{x'+vt'}{(1-\beta^2)^{1/2}} \qquad t=\frac{t'+\beta x'/c}{(1-\beta^2)^{1/2}}
\ee

Note that mutiplying (\ref{counting}) with (\ref{counting-}) we obtain the invariance of the interval between two events
\be
t^2-\left(\frac{x}{c}\right)^2 = t'^2-\left(\frac{x'}{c}\right)^2
\ee

\resection{The Lorentz Transformation for Arbitrary Velocities}

Usually one thinks of the STR as a theory that deals with inertial frames only, where acceleration has no place. This is not correct, and there are, indeed, transformation laws for the coordinates from a frame in arbitrary motion to a given inertial frame (the ``lab frame''). These were found by
Nelson in (\cite{Nelson}), and we derive here their one-dimensional version.

Consider two systems of coordinate in a similar way as done in section 9, where we derived the Lorentz transformations. The difference now is that Marianna's velocity can depend on time in an arbitrary way. At $t=t'=0$ their origins coincide, and they are receiving light pulses from a distant light source. Consider a physical event ${\cal E}$ that happens at some point in space-time. Let us count how many light pulses are there between the first pulse, that defined the $t=t'=0$ instant, and the pulse that crossed the event ${\cal E}$. According to Isabella, there are
\be
N_I=\left( t + \frac{x}{c} \right) \nu_0 \ . \label{NI}
\ee
To perform Marianna's counting we have to move along with her. Between $\tau$ and $\tau + d\tau$ she observes
\be
dN = d \tau \left( \frac{1+\beta}{1-\beta}\right)^{1/2} \nu_0 \label{d_pulses}
\ee
where now $\beta = \beta(\tau)$. She records the spacetime coordinates of the event ${\cal E}$ as being $(x',t')$, and therefore, between the first pulse and the pulse that defines the event, there are $x'/\lambda = x' \nu(t')/c$. Integrating (\ref{d_pulses}) from $0$ to $t'$ and adding $x' \nu(t')/c$ we get the total number of pulses
\be
N_M = \frac{x'}{c}\left(\frac{1+\beta(t')}{1-\beta(t')}\right)^{1/2} \nu_0+ \int_0^{t'} d\tau \left(\frac{1+\beta}{1-\beta}\right)^{1/2} \nu_0 \label{NM}
\ee 
Equating (\ref{NI}) and (\ref{NM}), and after some little algebra, we obtain
\be
t + \frac{x}{c} = \gamma(t') \frac{x'}{c} + \gamma(t') \beta(t') \frac{x'}{c} + \int_0^{t'} d\tau \gamma(\tau) + \int_0^{t'} d\tau \gamma(\tau) \beta(\tau) \label{t+x/c}
\ee
where we introduced $\gamma(\tau)=(1-\beta^2(\tau))^{-1/2}$. Like in the case of the Lorentz transformation, we still need one more equation. This is easily done by observing that in Isabella's frame, between $t$ and $t+dt$, Marianna has traveled $v(t)dt$, this will correspond to $\gamma(\tau)c\beta(\tau)d\tau$ in Marianna's coordinates, and the final portion, which according to Marianna, is $x'$, will be $\gamma(t')x'$, which gives
\be
x=\gamma(t') x' +\int_0^{t'} d\tau \gamma(\tau) c \beta(\tau) \label{x} \label{nelson_x}
\ee 
Substituting (\ref{x}) in (\ref{t+x/c}), we obtain, finally
\be
t = \gamma(t') \beta(t') \frac{x'}{c} + \int_0^{t'} d\tau \gamma(\tau) \label{nelson_t}
\ee
It is not easy to invert equations (\ref{nelson_x}) and (\ref{nelson_t}), except in a few cases, such as in the hyperbolic motion.
It is straightforward to see that these transformations reduce to Lorentz transformation in the case $\beta$ is constant.

It is instructive to apply these transformations to the case of hyperbolic motion, and use it to solve Bell's paradox. Before doing that, it is convenient to parametrize the velocity as $v=c \tanh(\phi)$, where $\phi$ is an arbitrary function of the proper time. The Lorentz transformation becomes
\bea
&&x=\cosh(\phi)x'+c\int_0^{t'} d\tau \sinh(\phi) \nonumber\\
&&c t=\sinh(\phi)x'+\int_0^{t'} d\tau \cosh(\phi) \label{LTphi}
\eea
The uniform acceleration problem is instructive. In this case $\phi = a \tau/c$, where $a$ is the acceleration measured in a frame of reference where the spaceship is instantenously at rest.

In general one approaches this problem by finding the coordinate $x'$, in $I´$'s frame, of the $M$ spaceship, in order to see that it is, indeed, pulling apart. We will do the opposite here: given that $M$ is at a certain spacetime point, which characterizes a physical event ${\cal E}$, with coordinates $(x,t)$ according to the lab frame, what are it's coordinates in $I$'s frame? As we will see, the computations are much simpler in this case.

In the lab we have, for the spaceship $M$,
\be
\left\{\begin{array}{l}
x_M= \alpha c(\cosh(\tau/\alpha)-1) + L_0\\
c t_M= \alpha c \sinh(\tau/\alpha)
\end{array}\right. \label{M}
\ee
The coordinates of a physical event are, in $I$'s coordinates, $(x',\tau')$, which are connected to the lab coordinates by
\be
\left\{\begin{array}{l}
x= (x'+\alpha c)\cosh(\tau'/\alpha)-\alpha c \\
c t= (x'+ \alpha c) \sinh(\tau'/\alpha)
\end{array}
\right. \label{I}
\ee
It is easy to see that, from (\ref{M}) and (\ref{I}) we obtain
\bea
&&(x'+\alpha c)^2 = (x_M+\alpha c)^2 -(ct_M)^2 = \alpha^2c^2+L_0^2+2\alpha c L_0 \cosh(\tau/\alpha) \nonumber \\
&&\sinh(\tau'/\alpha)=\frac{\alpha c \sinh(\tau/\alpha)}{(\alpha^2c^2+L_0^2+2\alpha c L_0 \cosh(\tau/\alpha))^{1/2}}
\eea
from which we can extract the asymptotic behaviour  $\exp(\tau'/\alpha) \approx (\alpha c/L_0)^{1/2} \exp(\tau/2\alpha) $ and $x' \approx (\alpha c L_0)^{1/2} \exp(\tau/2\alpha) \approx L_0 \exp(\tau'/2\alpha)$. We see, then, that the spaceship $M$ distances itself from $I$, which is, essentially due to the failure of simultaneity.

\resection{The Mass Energy Relation}

Finally, let us derive the mass-energy relation. Consider a body of mass $M$ that emits two bundles of radiation in opposite directions, each bundle carries an energy equal to $\delta E/2$. By symmetry the body will stay put. Before we proceed, let us make an observation concerning the linear momentum of the body and of the radiation.

We will assume that momentum and energy are conserved, and that the momentum of the body is given by $P=\xi(v/c)Mv$, based on dimensional grounds. Moreover, noticing that as $v \to -v$ we should have $P \to -P$, we deduce that $\xi(v/c) = \xi(-v/c)$. Since as $v \to 0$, $P=Mv$, we also have $\xi(0)=1$. These conditions imply that $\xi(v/c) = 1+\alpha (v/c)^2+...$, so that, whatever the relativistic momentum is, it equals $Mv$ up to factors of third order in $v/c$. Since in the following we will be looking at factors up to second order only, we may use $P=Mv$.

We will also be using the fact that the linear momentum carried by a bundle of radiation of energy $\delta E$ is $\delta E/c$. This relation preceeds the mass-energy relation and is usually derived \cite{Jackson} using conservation of total momentum (mechanical and electromagnetic) but does not assume an explicit form for the linear momentum.\footnote{Strictly speaking, we should remind the reader that in these derivations one uses the Lorentz force expression $d \vec p/dt = q(\vec E + v \times \vec B)$, which is relativistic, not manifestly, though.}

\begin{figure}[htbp]
	\centering
		\includegraphics{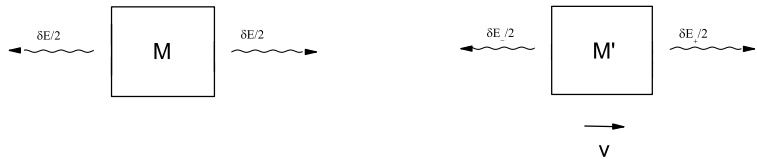}
	\caption{The same process as seen from two different reference frames.}
	\label{fig:2bodies}
\end{figure}

Let us analyse this is a frame that moves with velocity $-v$ along the same line as the emitted radiation as shown in figure \ref{fig:2bodies}.
In this frame, the energy of each bundle will change, and, since the body was at rest in the original frame, it will be moving with velocity $v$ in the second frame. If its mass is $M'$ now, then the linear momentum of the body is $M' v$.

In order to account the momentum carried by the radiation we have to find the expression for the energy of the radiation bundles in the ``moving" frame. In principle we could use the quantum relation that states that to a photon of energy $E$ corresponds the frequency $\nu=E/h$, and, by applying the Doppler formula, find $E'=E((1+\beta)/(1-\beta))^{1/2}$. This is correct, of course, but has one aspect that is less than satisfactory: the special theory of relativity is a completely classical theory, in the sense that its starting point is Maxwell's equations, and therefore we should not resort to any quantum relations in order to derive its results. Moreover, it is not obvious at all that the two theories could be put together in a simple and as direct manner as it is done in general, after all, why should these two theories be compatible? Therefore, instead of using quantum mechanics, we will take a somewhat longer route, but which is more elementary, and has the advantage of not appealing to quantum relations.

Using dimensional analysis, we can write an expression for the transformation of the energy of a bundle of radiation from one inertial frame to another, that moves with velocity $u$ in relation to it: it is simply $E'=Ef(\beta)$, where $f(\beta)$ is an unknown universal function, and $\beta=u/c$. This function must satisfy a composition law. Consider three inertial frames, $S_1$, $S_2$ and $S_3$, such that $S_2$ moves with velocity $v$ with relation to $S_1$, and $S_3$ moves with velocity $u$ with relation to $S_2$. Suppose there is a bundle of radiation in $S_1$ with energy $E_1$, then we must have $E_3=E_2f(u/c)=E_1f(v/c)f(u/c)$ and $E_3=E_1f((u/c+v/c)/(1+uv/c^2))$, where $E_i$ is the energy of the radiation measured in $S_i$, and we have used the formula for the addition of velocities. This means that:
\be
f(u/c)f(v/c)=f\left(\frac{u/c+v/c}{1+uv/c^2}\right)
\ee
It is not difficult to solve this equation, and we find (see appendix 2) that
\be
f(u/c)=\left(\frac{1+u/c}{1-u/c}\right)^{a/2}
\ee
where $a$ is a constant. Later we will show that $a=1$.

Proceeding with our argument, we can compute the momentum carried by light in the moving frame. Denoting by $\delta E_\pm$ the pulses that move to the right and to the left, we obtain
\be
\delta E_\pm = \frac{\delta E}{2}\left(\frac{1\pm \beta}{1\mp \beta}\right)^{a/2}
\ee
Therefore, momentum conservation gives
\be
M  v = M' v + \frac{\delta E}{2c}\left(\frac{1 + \beta}{1 - \beta}\right)^{a/2} - \frac{\delta E}{2c}\left(\frac{1 - \beta}{1 + \beta}\right)^{a/2}
\ee
This equation implies, up to second order in $v/c$ that
\be
(M'-M) v = \delta E a \frac{v}{c^2} \Rightarrow (M'-M) = a \frac{E}{c^2} \label{ae=mc^2}
\ee
This is not quite the final result, since we still need to fix $a$, which will done by using energy conservation. Once again, we can write the total energy of the body as $U(M,v)=\eta(v/c) Mc^2$ based on dimensional grounds. The function $\eta(v/c)$ should be such that, to second order in $v/c$, the difference in energy of a body of mass $M$ at rest and the same body with velocity $v$ should be the kinectic energy, that is $\eta(v/c)-\eta(0)= (v/c)^2/2$. And since the energy should not depend on the direction of the velocity, $\eta(v/c)=\eta(-v/c)$, and therefore $\eta(v/c)=\eta_0+(v/c)^2/2+o((v/c)^4)$.

Let us look at the energy conservation in both frames of reference. In the lab frame we have
\be
U(M,0)=U(M',0)+\delta E \label{energy1}
\ee
whereas in the moving frame we have, up to second order in $(v/c)^2$,
\be
U(M,v)=U(M',v)+\delta E_++\delta E_- \Rightarrow U(M,0)+\frac{Mv^2}{2}=U(M',0)+\frac{M'v^2}{2}+\delta E (1+a^2\beta^2)
\label{energy2}
\ee
Equations (\ref{energy1}) and (\ref{energy2}) imply that $(M-M')v^2/2=a^2 \delta E \beta^2$, which together with (\ref{ae=mc^2}), fixes $a=1$, and establishes the well-known mass-energy relation
\be
\delta E = (M-M')c^2
\ee
Before closing this section, let us find the exact forms of the functions $\xi(v/c)$ and $\eta(v/c)$. Let us find $\eta(v/c)$ first.

Suppose the body emits two bundles of radiation with the same energy, to the left and right, each with energy $\delta E$. After this emission the mass of the body is $M'=M-\delta E/c^2$. In the lab frame we have from energy conservation, $U(M,0)=U(M',0)+ \delta E/2+ \delta E/2$, while in the moving frame we have $U(M,v)=U(M',v)+\delta E_+ + \delta E_-$, and therefore
\be
\eta(v/c)Mc^2=\eta(v/c)M'c^2+
\frac{\delta E}{2}\left(\left(\frac{1+\beta}{1-\beta}\right)^{1/2}+
\left(\frac{1-\beta}{1+\beta}\right)^{1/2}\right)=\frac{\delta E}{(1-\beta^2)^{1/2}} = \gamma \delta M c^2
\ee
which implies that $\eta(v/c)=\gamma$, and the energy is given by $U(M,v)=\gamma M c^2$.

Finally, for momentum we have, following a setup similar to the one just described, before the emission of radiation, $P=\xi(v/c)Mv$, and after $\xi(v/c)M'v+\delta E_+/c-\delta E_-/c$, which gives
\bea
&&\xi(v/c)Mv =\xi(v/c)M'v + \frac{\delta E}{2}\left(\left(\frac{1+\beta}{1-\beta}\right)^{1/2}-
\left(\frac{1-\beta}{1+\beta}\right)^{1/2}\right)= \nonumber \\ 
&&\phantom{\xi(v/c)Mv=}\xi(v/c)M'v+\frac{\delta Ev/c^2}{(1-\beta^2)^{1/2}}=\xi(v/c)M'v+\gamma \delta M v
\eea
which fixes the function $\xi(v/c)=\gamma$. Note that we can write these expressions in terms of the derivatives of $t$ and $x$ with respect to the proper time of the particle $\tau$, since $dt/d\tau = \gamma$. We are then led naturally to the definition of the `four-momentum' (in this case the `two-momentum') $P^\mu=(P^0,P^1)$, with $P^0=U/c=Md(ct)/d\tau$ and $P^1=P=Mdx/d\tau$, which are expressed in terms of the four-velocity by $P^\mu=Mdx^\mu/d\tau$. 

These two relations can be used as the starting point to the study of relativistic dynamics. This completes our discussion of special relativity.

\resection{Conclusions}

We have derived all relativistic kinematics, taking the Doppler effect as the starting point. The novelty in this presentation is that we derive the Doppler effect {\em directly} from the relativity postulates, and from there obtain everything else: time dilation, length contraction, addition of velocities, Lorentz transformations for constant velocity and for arbitrary velocities, the mass-energy relation, and the introduction of relativistic energy and momentum for a massive body. We also used it to explain the ``twin paradox". Our approach is particularly useful in the discussion of the twin paradox on a cylinder. In deriving the mass-energy relation we used only fairly general arguments, staying always in the realm of elementary classical physics: we did not use the energy/frequency relation for photons from quantum mechanics, nor used the classical expression of a wave packet in terms of electromagnetic fields, to derive the fact that the energy will also be Doppler shifted. We should stress the fact that, in deriving the relativistc energy and momentum, we assumed that there are conserved quantities like energy and momentum, and that light interacts with matter in such a way that all energy can be absorbed or emitted. 

Another feature of this approach is that all arguments presented are truly one-dimensional: we did not use light rays moving perpendicularly to its motion, for example, as is done in some elementary presentations of the special theory of relativity. 

\resection{Acknowledgments}
I would like to thank L. Moriconi for several useful discussions, a critical reading of this note, and for insisting in making the presentation of the mass-energy relation elementary, helping along the way with several crucial insights. A very nice email exchange with N. D. Mermin is gratefully acknowledged, as well as his sending me chapter 7 of his forthcoming book (\cite{Nelson}), which has some overlap with the arguments presented in this paper, especially the derivation of the Doppler shift expression and the addition of velocities. Thanks are also due to M. Parikh, for telling me about the twin paradox on a cylinder, and N. Lemos, for a critical reading of this paper. This work has been partially supported by Faperj.

\newpage

\section*{Appendix 1}
\renewcommand{\theequation}{1.\arabic{equation}}
  % redefine the command that creates the equation no.
  \setcounter{equation}{0}
  
In this appendix we present another elementary derivation of the Doppler effect. Suppose that Isabella is in an inertial frame of reference, and that she is sending light pulses at a frequency $\nu_0$, and that Marianna is moving towards her with a mirror. As explained in the paper, she will observe pulses at frequency $\nu_1=f(\beta)\nu_0$, and since her mirror reflects them, this is equivalent to her carrying a light source sending pulses at frequency $\nu_1$. Finally, Isabella will observe pulses at a frequency $\nu_2=f(\beta)\nu_1=f(\beta)^2\nu_0$. 

Let us analyze two sequential pulses reflecting from Marianna´s mirror, call them ``pulse-1" and ``pulse-2", as seen in Isabella´s frame. Let the distance between these two pulses be $\lambda_0$. It is easy to see that pulse-2 hits the mirror a time $\Delta t$ after pulse-1, given by
\be
c \Delta t + v \Delta t = \lambda_0 \ .
\ee
During this time interval, pulse-1 has traveled $c\Delta t$ and the mirror has traveled $v \Delta t$, therefore the distance between the two pulses, which is the wave length measured by Isabella, is given by
\be
\lambda = (c-v) \Delta t \ .
\ee
Now we use the constancy of the speed of light: we know that $\lambda_0 \nu_0 = c$ and that $\lambda \nu_2 = c$, which imply
\be
(c-v) \Delta t \nu = (c+v)\Delta t \nu_0 \Rightarrow \nu = \frac{c+v}{c-v} \nu_0
\ee
and therefore
\be
f(\beta) = \left( \frac{1+\beta}{1-\beta}\right)^{1/2}
\ee

\newpage

\section*{Appendix 2}
\renewcommand{\theequation}{2.\arabic{equation}}
  % redefine the command that creates the equation no.
  \setcounter{equation}{0}
  
We want to solve the functional equation
\be
f(u/c)f(v/c)=f\left(\frac{u+v}{1+uv/c^2}\right) \label{ff}
\ee
Let us take $u/c$ arbitrary and $v/c$ infinitesimal, equal to $\epsilon$. Expanding both sides of (\ref{ff}) in Taylor series, we obtain
\be
f(u/c)(f(0)+\epsilon a +\ldots)=f((u/c+\epsilon)(1-\epsilon u/c^2))=f(u/c)+\epsilon f'(u/c)(1-u^2/c^2)+\ldots
\ee 
where we introduced $a=f'(0)$, and can set $f(0)=1$ for physical reasons: if the observer is at rest with the light source, then the energy measured by the observer should be the same as the one emitted. This expansion gives us the elementary differential equation
\be
f'(u/c)=\frac{a f(u/c)}{1-u^2/c^2}
\ee
This can be easily solved, and we obtain
\be
f(u/c)=\left(\frac{1+u/c}{1-u/c}\right)^{a/2} \label{solution}
\ee
If the reader feels this is a little too involved, or would rather avoid the use of calculus, there is another derivation we can provide \footnote{I thank N. Mermin for this suggestion.}. Replace $v/c$ by $\tanh(x)$, and $u/c$ by $\tanh(y)$, and define $g(x)= \ln f(\tanh(x))$. Equation (\ref{ff}) becomes
\be
g(x)+g(y)=g(x+y)
\ee
whose solution is trivially seen to be $g(x)=ax$, or, equivalently, $f(\tanh(x))=ax$, which is easy to show to be the same as (\ref{solution}). 
\newpage

\section*{Appendix 3}
\renewcommand{\theequation}{3.\arabic{equation}}
  % redefine the command that creates the equation no.
  \setcounter{equation}{0}

This is an alternative derivation of the Lorentz transformation using time dilation and length contraction. Its method  does lie outside the main line of reasoning of the paper, but we found it could be useful to have an alternative way to derive the Lorentz transformation, using time dilation and length contraction as the starting point. 

Initially, let us call the laboratory frame $S$, and an inertial frame moving with velocity $v$ along the $x$ direction, $S_0$. We want to find the transformation law of the coordinates $(x_0,t_0)$ of a given physical ``event" $E_0$ in $S_0$, to the coordinates $(x,t)$ of this same event, as observed in $S$ (and which we refer to as $E$). We assume that $(x_0,t_0)=(0,0)$ corresponds to $(x,t)=(0,0)$.

In order to find this transformation law, let us assume that at the same time and position of $E_0$, a light flash was emitted. This light flash will arrive at the origin of $S_0$ at time $t_0+x_0/c$. Due to the time dilation formula, this corresponds to time $\gamma (t_0+x_0/c)$ as measured in $S$, where $\gamma^{-1}=(1-\beta^2)^{\frac{1}{2}}$. In order to find the time the event $E$ took place when measured in $S$, we have to ``roll the film backwards". Since the light arrived at the origin of $S_0$ at time (measured in $S$) $\gamma (t_0+x_0/c)$, it means that it left the event $E$ at time $\gamma (t_0+x_0/c) - \gamma^{-1} x_0/(c+v)$: we have to subtract the time light takes to cover the distance $\gamma^{-1} x_0$ to the origin, as the origin moves with velocity $v$. This gives:
\be
t=\gamma(t_0+\frac{x_0}{c})-\gamma^{-1}\frac{x_0}{c+v} \Rightarrow
t=\frac{t_0+\beta x_0/c}{(1-\beta^2)^{1/2}} \label{time}
\ee
We can perform a similar argument to obtain the transformation law for the spatial coordinate. At the time light from the event $E_0$ arrives at the origin of $S_0$, the origin itself has traveled a distance $v\gamma^{-1}(t_0+x_0/c)$, according to $S$. We have to subtract from this value the distance traveled by the origin of $S_0$ {\em during} the time light is going from $E$ to the origin of $S_0$, which is given by $v \gamma^{-1} x_0/(c+v)$, and add the length contracted distance $\gamma^{-1}x_0$. This gives
\be
x=v\gamma (t+\frac{x_0}{c})-v\gamma^{-1}\frac{x_0}{c+v}+\gamma^{-1}x_0 \Rightarrow 
x=\frac{x_0+vt_0}{(1-\beta^2)^{1/2}}\label{position}
\ee
Equations (\ref{position}) and (\ref{time}) are the Lorentz transformation of coordinates for one dimensional relativistic motion.

\end{document}